\documentclass[12pt]{article}%
\usepackage{amsmath}
\usepackage{amsfonts}
\usepackage{amssymb}
\usepackage{graphicx}%
\begin{document}
\begin{titlepage}
\vspace{.3cm} \vspace{1cm}
\begin{center}
\baselineskip=16pt \centerline{\Large\bf Scale Invariance in the
Spectral Action } \vspace{2truecm} \centerline{\large\bf Ali H.
Chamseddine$^1$\ , \ Alain Connes$^{2,3,4}$\ \ } \vspace{.5truecm}
\emph{\centerline{$^1$Center for Advanced Mathematical Sciences and}
\centerline{Physics Department, American University of Beirut,
Lebanon.} }
\end{center}
\emph{\centerline{$^2$  College de France, 3 rue Ulm, F-75005,
Paris, France} \centerline{$^3$ I.H.E.S. Bures-sur-Yvette, France},
\centerline{$^4$ Vanderbilt University, Tennessee USA.}} \vfill
\begin{center}
{\bf Abstract}
\end{center}
The arbitrary mass scale in the spectral action for the Dirac
operator in the spectral action is made dynamical  by introducing a
dilaton field. We evaluate all the low-energy terms in the spectral
action and determine the dilaton couplings. These results are
applied to the spectral action of the noncommutative space defined
by the standard model. We show that the effective action for all
matter couplings is scale invariant, except for the dilaton kinetic
term and Einstein-Hilbert term. The resulting action is almost
identical to the one proposed for making the standard model scale
invariant as well as the model for extended inflation and has the
same low-energy limit as the Randall-Sundrum model. Remarkably, all
desirable features with correct signs for the relevant terms are
obtained uniquely and without any fine tuning.
\end{titlepage}

\section{\bigskip Introduction}

\bigskip It is known that the standard model of strong and electroweak
interactions is classically almost scale invariant, and that the only terms
that break the dilatation symmetry are the mass terms in the Higgs sector.
\ Scale invariance of the classical Lagrangian can be achieved by introducing
a compensating dilaton field \cite{Coleman}, \cite{Bud}, \cite{Bubu}. Breaking
of scale invariance occurs after the electroweak symmetry is broken
spontaneously through the generation of radiative corrections to the scalar
potential. The dilaton mass scale is much larger than the weak scale and could
be as large as the GUT scale or Planck scale. This leads naturally to consider
the coupling of the dilaton to gravity. The dilaton is always part of the low
energy spectrum in string theory. Historically it first appeared in the
Jordan-Brans-Dicke theory of gravity which corresponds to one particular
coupling of the dilaton to the metric \cite{Bud}. The dilaton plays a
fundamental role in models of inflation \cite{Kolb}. It also appears in the
gravitational couplings of the noncommutative Connes-Lott formulation of the
standard model \cite{CL}, \cite{CFF},\cite{CFG}, \cite{CF}, where the dilaton
is the scalar field that couples the two sheets of space-time. The resulting
matter interactions in this case are also scale invariant, and the
gravitational couplings are different than the Jordan-Brans-Dicke theory. More
recently, a scalar field, the radion field, appeared in the Randall-Sundrum
scenario (RS) of compactification \cite{RS} which is related to the question
of masses and scales in physics. The RS scenario was shown to be equivalent to
the results derived from noncommutative geometry, \cite{Lizzi}, which is not
too surprising, because both the Connes-Lott model and the RS model, describe
a system with two branes.

At present, and within the noncommutative geometric picture, the spectral
action gives the most elegant formulation of the standard model \cite{ACAC},
\cite{Kastler}. All details of the standard model as well as its unification
with gravity are achieved by postulating the action
\[
\mathrm{Trace}\,F\left(  D^{2}/m^{2}\right)  +\left\langle \Psi\left\vert
D\right\vert \Psi\right\rangle ,
\]
where $D$ \ is the Dirac operator of a certain noncommutative space and $\Psi$
is a spinor in the Hilbert space of the observed quarks and leptons. However,
the dilaton field does not appear in the spectral action, which is to be
contrasted with the Connes-Lott formulation of the noncommutative action where
the dilaton field is part of the gravitational interactions. This suggests
that the Dirac operator used in the construction of the spectral action should
be modified in order to take into account the presence of the dilaton.

The appearance of the dilaton field in physical models is related to the
question of mass and scales. It is therefore natural to consider replacing the
mass parameter in the ratio $D^{2}/m^{2}$ by a function of the dilaton, thus
introducing a sliding scale, as the Dirac operator have the dimension of mass.
This is also relevant when dealing with non-compact manifolds where $D$ no
longer has discrete spectrum and the counting of eigenvalues requires a
localization. \ Let the dynamical scale factor $\rho$ be written in the form
\[
\rho=me^{\phi},
\]
where we assume that $\phi$ is dimensionless. The dilaton can be related to a
scalar field $\sigma$ of dimension one by writing
\[
\phi=\frac{1}{f}\sigma,
\]
where $f$ \ is the dilaton decay constant. The mass scale $m$ can be absorbed
by the redefinition%
\[
\phi\rightarrow\phi-\ln m,
\]
and therefore we can assume, without any loss in generality, that
$\rho=e^{\phi}.$ One can always recover the scale $m$ by performing the
opposite transformation $\phi\rightarrow\phi+\ln m$. Now using $\rho$ instead
of the scalar $m$ in the counting of eigenvalues :
\[
N(m)=\,\mathrm{Dim}\,\{ D^{2} \leq m^{2}\} \to N(\rho)=\,\mathrm{Dim}\,\{
D^{2} \leq\rho^{2}\}
\]
is equivalent to replacing the operator $D^{2}/m^{2}$ in \cite{ACAC} by
\[
P=e^{-\phi}D^{2}e^{-\phi}\,.
\]
If we insist that the metric $g_{\mu\nu}$ be dimensionless to insure that its
flat limit be the Minkowski metric, then the scale $m$ \ will explicitly
appear in the action after rescaling $e^{\phi}\rightarrow me^{\phi}$.
Otherwise we can absorb this mass scale by assuming that the metric has the
dimension of mass.

The aim of this paper is to determine the interactions of the dilaton field
$\phi$ with all other fields present in the spectral action formulation of the
standard model. Because of the spectral character of the action, it is
completely determined from the form of \ $P$ and there is no room for fine
tuning the results. It is then very reassuring to find that the resulting
interactions are identical to those constructed in the literature by
postulating a hidden scale invariance of the matter interactions \cite{Bubu}.
These are also equivalent to the interactions of the radion field in the RS
model \cite{RS}. All of these results now support the conclusion that
space-time at high-energies reveals its discrete structure, and is governed by
noncommutative geometry. The plan of this paper is as follows. In section two
we briefly review the derivation of the spectral action and comment on the
modifications needed to include the dilaton. In section three we derive the
Seeley-de Witt coefficients of the spectral action in presence of the dilaton.
In section four we give the full low-energy spectral action including dilaton
interactions specialized to the noncommutative space of the standard model. In
section five we compare our results with those obtained by imposing scale
invariance on the standard model interactions, to the RS model and the model
of extended inflation. Section six is the conclusion. The appendices contain
detailed proofs of some identities used.

\section{\bigskip A summary of spectral action}

We begin by summarizing the results of \cite{ACAC}. The square of the Dirac
operator appearing in the spectral triple of a noncommutative space is written
in the following form suitable to apply the standard local formulas for the
heat expansion (see \cite{Gilkey} section 4.8).
\begin{equation}
D^{2}=-\left(  g^{\mu\nu}I\,\partial_{\mu}\partial_{\nu}+A^{\mu}\partial_{\mu
}+B\right)  ,\label{opdata}%
\end{equation}
where $g^{\mu\nu}$ plays the role of the inverse metric, $I$ is the unit
matrix, $A^{\mu}$ and $B$ are matrix functions computed from the Dirac
operator. The bosonic part of the spectral action can be expanded in a power
series as function of the inverse scale $m,$ and is given in dimension $4$ by
\[
\mathrm{Trace}\left(  F\left(  D^{2}/m^{2}\right)  \right)  \simeq%
{\displaystyle\sum\limits_{n\geq0}}
\,f_{n}\,a_{n}\left(  D^{2}/m^{2}\right)  ,
\]
where $F$ is a positive function and
\begin{align}
f_{0} &  =%
{\displaystyle\int\limits_{0}^{\infty}}
F\left(  u\right)  udu,\quad f_{2}=%
{\displaystyle\int\limits_{0}^{\infty}}
F\left(  u\right)  du,\quad\label{fncoeffs}\\
\qquad f_{2n+4} &  =\left(  -1\right)  ^{n}F^{\left(  n\right)  }\left(
0\right)  \,,\quad n\geq0\,.\nonumber
\end{align}
The positivity of the function $F$ will insure that the actions for gravity,
Yang-Mills, Higgs couplings are all positive and the Higgs mass term is
negative. We will comment on the positive sign of the cosmological constant at
the end of the paper.

The first few Seeley-deWitt coefficients $a_{n}\left(  D^{2}/m^{2}\right)  $
are given (see \cite{Gilkey} Theorem 4.8) by\footnote{According to our
notations the scalar curvature $R$ is negative for spheres (see \cite{Gilkey}
section 2.3) and the space is Euclidean.}
\begin{align}
a_{0}\left(  D^{2}/m^{2}\right)   &  =\frac{m^{4}}{16\pi^{2}}%
{\displaystyle\int\limits_{M}}
d^{4}x\sqrt{g}\,Tr(1),\\
a_{2}\left(  D^{2}/m^{2}\right)   &  =\frac{m^{2}}{16\pi^{2}}%
{\displaystyle\int\limits_{M}}
d^{4}x\sqrt{g}\,Tr\left(  -\frac{R}{6}+E\right)  ,\\
a_{4}\left(  D^{2}/m^{2}\right)   &  =\frac{1}{16\pi^{2}}\frac{1}{360}%
{\displaystyle\int\limits_{M}}
d^{4}x\sqrt{g}\,Tr\left(  -12R_{;\mu}^{\;\;\mu}+5R^{2}-2R_{\mu\nu}R^{\mu\nu
}\right.  \\
&  \quad\left.  +\,2R_{\mu\nu\rho\sigma}R^{\mu\nu\rho\sigma}-60RE+180E^{2}%
+60E_{;\mu}^{\quad\mu}+30\,\Omega_{\mu\nu}\Omega^{\mu\nu}\right)  \nonumber
\end{align}
while the odd ones all vanish
\[
a_{2n+1}\left(  D^{2}/m^{2}\right)  =0.
\]

The notations are as follows, one lets $\Gamma_{\mu\nu}^{\rho}\left(
g\right)  $ be the Christoffel symbols of the Levi-Civita connection of the
metric $g$ and lets
\[
\Gamma^{\rho}\left(  g\right)  =g^{\mu\nu}\Gamma_{\mu\nu}^{\rho}\left(
g\right)
\]
The connection form $\overline{\omega}$, its curvature $\Omega,$ and the
endomorphism $E$ are then defined by (see \cite{Gilkey} section 4.8)
\begin{align}
\overline{\omega}_{\mu}  &  =\frac{1}{2}g_{\mu\nu}\left(  A^{\nu}+\Gamma^{\nu
}\left(  g\right)  I\right)  ,\label{notations}\\
\Omega_{\mu\nu}  &  =\partial_{\mu}\overline{\omega}_{\nu}-\partial_{\nu
}\overline{\omega}_{\mu}+\left[  \overline{\omega}_{\mu},\overline{\omega
}_{\nu}\right]  ,\\
E  &  =B-g^{\mu\nu}\left(  \partial_{\mu}\overline{\omega}_{\nu}%
+\overline{\omega}_{\mu}\overline{\omega}_{\nu}-\Gamma_{\mu\nu}^{\rho}\left(
g\right)  \overline{\omega}_{\rho}\right)  \,.
\end{align}
To understand algebraically the dependence in the operator $D$ it is
convenient to express the above coefficients as residues and this is done as
follows in the generality that we need. One lets $P$ be a second order
elliptic operator with positive scalar principal symbol and defines a zeta
function as
\[
\zeta_{P}(s)=\mathrm{Trace}\,(P^{-s/2})\label{eq1.70}%
\]
One then gets in the required generality for our purpose the equality
\begin{align}
\mathrm{Trace}\,(F(P/m^{2}))  &  \sim\frac{m^{4}}{2}\ \mathrm{Res}%
_{s=4}\,\zeta_{P}(s)f_{0}\,\nonumber\\
&  +\frac{m^{2}}{2}\ \mathrm{Res}_{s=2}\,\zeta_{P}(s)f_{2}\,+\zeta
_{P}(0)\,f_{4}+\cdots.\nonumber
\end{align}
which, using the Wodzicki residue which is given independently of $D$ by
\begin{equation}
\int\!\!\!\!\!\!-\ T=\ \mathrm{Res}_{s=0}^{{}}\,\mathrm{Trace}\,(T(D^{2}%
)^{-s/2}), \label{res}%
\end{equation}
can be written as
\begin{align}
\mathrm{Trace}\,(F(P/m^{2}))  &  \sim\frac{m^{4}}{2}\,f_{0}\,\int
\!\!\!\!\!\!-\ P^{-2}\label{resform}\\
&  +\frac{m^{2}}{2}\,f_{2}\,\int\!\!\!\!\!\!-\ P^{-1}\,+\,f_{4}\,\zeta
_{P}(0)+\cdots.\nonumber
\end{align}
We want to compute the spectral action associated with the operator
$P=e^{-\phi}D^{2}e^{-\phi}$ i. e. to determine the dependence of the spectral
action on the dilaton field $\phi$. The first term $\int\!\!\!\!\!\!-\ P^{-2}$
is a Dixmier trace and one can permute the functions with the operators
without altering the result since the Dixmier trace vanishes on operators of
order $<-4$. One thus gets, for any test function $h$,
\[
\int\!\!\!\!\!\!-\ h\,P^{-2}=\,\int\!\!\!\!\!\!-\ h\,e^{4\phi}\,D^{-4}%
\]
which shows that we get an overall factor of $e^{4\phi}$ multiplying
$a_{0}\left(  x,D^{2}\right)  $. For the second term $\int
\!\!\!\!\!\!-\ P^{-1}$ we have
\[
\int\!\!\!\!\!\!-\ h\,P^{-1}=\,\int\!\!\!\!\!\!-\ h\,e^{\phi}D^{-2}e^{\phi
}=\,\int\!\!\!\!\!\!-\ h\,e^{2\phi}D^{-2}%
\]
using the trace property of the residue and again we get an overall factor of
$e^{2\phi}$ multiplying $a_{2}\left(  x,D^{2}\right)  $. Note that the result
remains valid when the test function $h$ is taken with values in endomorphisms
of the vector bundle on which $P$ is acting. This suggests that the invariance
of the $a_{2}$ term (up to the $e^{2\phi}$ scale factor) takes place before
taking the fiberwise trace. The direct computation below in equation
\eqref{invscale} will confirm this point .

The term $f_{4}\,\zeta_{P}(0)$ is more tricky to analyze and we shall only
give now a heuristic argument explaining why it should be independent of
$\phi$. We shall then check it by a direct calculation. The formal argument
proceeds as follows. First one lets
\[
P(t)=\,e^{-t\phi}\,D^{2}\,e^{-t\phi}%
\]
so that $P(0)=D^{2}$ and $P(1)=P$ with the above notations. Let then
\[
Y(t)=\,\mathrm{Log}\,P(t)-\,\mathrm{Log}\,P(0).
\]
Using the equality ( $a>0$ )%
\[
\mathrm{Log\;}a=%
{\displaystyle\int\limits_{0}^{\infty}}
\left(  \frac{1}{\lambda+1}-\frac{1}{\lambda+a}\right)  \,d\lambda,
\]
applied to $P(t)$ one obtains the relation
\begin{equation}
\frac{d}{dt}\,Y(t)=\,-\,{\displaystyle\int\limits_{0}^{\infty}}\,(P(t)+\lambda
)^{-1}\,(\phi\,P(t)+\,P(t)\,\phi)\,(P(t)+\lambda)^{-1}\,d\lambda. \label{derY}%
\end{equation}
One then shows that
\begin{equation}
\frac{d}{dt}\,Y(t)=\,-2\,\phi+\,[P(t),C(t)], \label{commutator}%
\end{equation}
where $C(t)$ is a pseudodifferential operator (see appendix A). Thus one gets
a similar expression
\begin{equation}
Y=\,Y(1)=\,-2\,\phi+\,{\displaystyle\int\limits_{0}^{1}}\,[P(t),C(t)]\,dt.
\label{commutator1}%
\end{equation}
Next one uses the expansional formula
\[
e^{A+B}\,e^{-A}=\sum_{0}^{\infty}\int_{0\leq t_{1}\leq\cdots\leq t_{n}\leq
1}B(t_{1})\,B(t_{2})\ldots B(t_{n})\,\prod\,dt_{i}%
\]
where
\[
B(t)=e^{tA}\,B\,e^{-tA}\,.
\]
One lets $A=\,-s\,\mathrm{Log}\,P(0)$ and $B=\,-s\,Y$. This gives an equality
of the form
\[
P^{-s}=\,D^{-2s}-s\,\int_{0}^{1}\sigma_{-st}(Y)\,dt\,D^{-2s}\,
\]%
\[
+\,\sum_{2}^{\infty}\,(-s)^{n}\,\int_{0\leq t_{1}\leq\cdots\leq t_{n}\leq
1}\sigma_{-st_{1}}(Y)\,\sigma_{-st_{2}}(Y)\ldots\sigma_{-st_{n}}%
(Y)\,\prod\,dt_{i}%
\]
where
\[
\sigma_{u}(T)=(D^{2})^{u}\,T\,(D^{2})^{-u}.
\]
One infers from this equality and the absence of poles of order $>1$ in the
zeta functions of the form $\mathrm{Tr}\,(Q\,D^{-2s})$ that the terms of order
$n>1$ in $s$ will not contribute to the value at $s=0$. Thus the following
should hold
\[
\zeta_{P}(0)-\,\zeta_{D^{2}}(0)=\,-\frac{1}{2}\,\int\!\!\!\!\!\!-\ Y
\]
and using \eqref{commutator1} one gets
\[
\zeta_{P}(0)-\,\zeta_{D^{2}}(0)=\,\int\!\!\!\!\!\!-\ \phi=\,0,
\]
since the residue vanishes on differential operators. It would take a lot more
care to really justify the above manipulations. Instead, in the next section,
we shall show by a brute force calculation that $a_{4}$ is independent of
$\phi$ so that the above identity is valid.

We thus see that in the first few terms of the spectral action, the only
modification we expect when the operator $D^{2}$ is replaced by $P$ \ is to
get an overall factor of $e^{\left(  4-n\right)  \phi}$ multiplying
$a_{n}\left(  x,D^{2}\right)  :$%
\[
\mathrm{Trace}(F(P))\backsimeq%
{\displaystyle\sum\limits_{n=0}^{6}}
f_{n}%
{\displaystyle\int}
d^{4}x\sqrt{g}\,e^{\left(  4-n\right)  \phi}a_{n}(x,D^{2})+\cdots
\]
Also as will be shown in appendix B, we have the identity
\[
a_{n}(x,e^{-\phi}D^{2}e^{-\phi})=a_{n}\left(  x,D^{2}e^{-2\phi}\right)
=a_{n}(x,e^{-2\phi}D^{2}).
\]
It is easy to check that by applying the inverse transformation $\phi
\rightarrow\phi+\ln m$ \ one recovers all the $m$ scaling factors obtained in
reference \cite{ACAC}. \ In the next section, we shall confirm this result by
directly evaluating the spectral action associated with the operator $P$ and
in particular the low-energy terms $a_{0},$ $a_{2}$ and $a_{4}.$ We will not
attempt to evaluate higher order terms as these are not needed in our analysis.

\section{\bigskip Dilaton and Seeley-deWitt coefficients}

\label{seeleydewitt}

We compare quite generally the Seeley-deWitt coefficients of an operator
$P_{0}=\,D^{2}$ given by \eqref{opdata} and those of the rescaled operator
$P=e^{-\phi}D^{2}e^{-\phi}$. We use the rescaled metric $G$ in the Einstein
frame, where the dilaton factor is absorbed in the metric. First we write%
\begin{equation}
\label{opdatanew}P=e^{-\phi}D^{2}e^{-\phi}=-\left(  G^{\mu\nu}\partial_{\mu
}\partial_{\nu}+\mathcal{A}^{\mu}\partial_{\mu}+\mathcal{B}\right)  ,
\end{equation}
where
\begin{align*}
G^{\mu\nu}  &  =e^{-2\phi}g^{\mu\nu},\\
\mathcal{A}^{\mu}  &  =e^{-2\phi}A^{\mu}-2G^{\mu\nu}\partial_{\nu}\phi,\\
\mathcal{B}  &  =e^{-2\phi}B+G^{\mu\nu}\left(  \partial_{\mu}\phi\partial
_{\nu}\phi-\partial_{\mu}\partial_{\nu}\phi\right)  -e^{-2\phi}A^{\mu}%
\partial_{\mu}\phi.
\end{align*}
The Seeley-deWitt coefficients for $a_{n}\left(  P\right)  $ are expressed in
terms of $\mathcal{E}$ and $\mathbf{\Omega}_{\mu\nu}$ defined by
\eqref{notations} so that,
\begin{align*}
\mathcal{E}  &  =\mathcal{B}-G^{\mu\nu}\left(  \partial_{\mu}\overline
{\omega^{\prime}}_{\nu}+\overline{\omega^{\prime}}_{\mu}\overline
{\,\omega^{\prime}}_{\nu}-\Gamma_{\mu\nu}^{\rho}\left(  G\right)
\overline{\omega^{\prime}}_{\rho}\right)  ,\\
\overline{\omega^{\prime}}_{\mu}  &  =\frac{1}{2}G_{\mu\nu}\left(
\mathcal{A}^{\nu}+\Gamma^{\nu}\left(  G\right)  \right)  ,\\
\mathbf{\Omega}_{\mu\nu}  &  =\partial_{\mu}\overline{\omega^{\prime}}_{\nu
}-\partial_{\nu}\overline{\omega^{\prime}}_{\mu}+\left[  \overline
{\omega^{\prime}}_{\mu},\overline{\omega^{\prime}}_{\nu}\right]  .
\end{align*}
These relations imply that
\[
\overline{\omega^{\prime}}_{\mu}=\frac{1}{2}g_{\mu\nu}A^{\nu}-\partial_{\mu
}\phi+\frac{1}{2}G_{\mu\nu}\Gamma^{\nu}\left(  G\right)  .
\]
The conformal transformations of the Christoffel connection give%
\begin{align*}
\Gamma_{\nu\rho}^{\mu}\left(  G\right)   &  =\Gamma_{\nu\rho}^{\mu}\left(
g\right)  +\left(  \delta_{\nu}^{\mu}\partial_{\rho}\phi+\delta_{\rho}^{\mu
}\partial_{\nu}\phi-g_{\nu\rho}g^{\mu\sigma}\partial_{\sigma}\phi\right)  ,\\
\Gamma^{\mu}\left(  G\right)   &  =e^{-2\phi}\Gamma^{\mu}\left(  g\right)
-2e^{-2\phi}g^{\mu\nu}\partial_{\nu}\phi.
\end{align*}
Using these relations we finally get
\begin{align}
\label{fundrel}\overline{\omega^{\prime}}_{\mu}  &  =\overline{\omega}_{\mu
}-2\partial_{\mu}\phi,\\
\mathcal{E}  &  =e^{-2\phi}\left(  E+g^{\mu\nu}\left(  \nabla_{\mu}^{g}%
\nabla_{\nu}^{g}\phi+\partial_{\mu}\phi\partial_{\nu}\phi\right)  \right)  ,
\label{fundrel1}%
\end{align}
where the covariant derivative $\nabla_{\mu}^{g}$ is taken with respect to the
metric $g.$ It is quite striking that the perturbation is only a scalar
multiple of the identity matrix and does not involve the endomorphisms
$A^{\mu}$ at all.

The term $a_{0}$ only involves $\sqrt{G} \,\mathrm{Tr}(1)$ which, when
expressed in terms of the metric $g$ \ gives $\sqrt{g}e^{4\phi}\,\mathrm{Tr}%
(1)$.

The $a_{2}$ term is proportional to
\[
\int d^{4}x\sqrt{G}\,Tr\left(  \mathcal{E}-\frac{1}{6}R\left(  G\right)
\right)  ,
\]
where the curvature scalar is constructed as function of the metric $G.$ We
now use \eqref{fundrel1} and the relation
\begin{equation}
\label{scalar}R\left(  G\right)  =e^{-2\phi}\left(  R\left(  g\right)
+6g^{\mu\nu}\left(  \nabla_{\mu}^{g}\nabla_{\nu}^{g}\phi+\partial_{\mu}%
\phi\partial_{\nu}\phi\right)  \right)
\end{equation}
to obtain at the level of endomorphisms (before taking the fiberwise trace)
\begin{equation}
\label{invscale}\mathcal{E}-\frac{1}{6}R\left(  G\right)  =e^{-2\phi}\left(
E-\frac{1}{6}R\left(  g\right)  \right)  .
\end{equation}
This of course implies the required rescaling of the $a_{2}$ term in the
required generality, but it is more precise since it holds before taking the
trace. We shall use this more precise form in the proof of the invariance of
the $a_{4}$ term.

The term $a_{4}\left(  P\right)  $ is given by
\begin{align*}
&  \frac{1}{16\pi^{2}}\frac{1}{360}\int d^{4}x\sqrt{G}\,Tr\left[  \left(
5R^{2}\left(  G\right)  -2R_{\mu\nu}(G)R^{\mu\nu}\left(  G\right)  +2R_{\mu
\nu\rho\sigma}\left(  G\right)  R^{\mu\nu\rho\sigma}(G)\right)  \right. \\
&  \hspace{1in}\hspace{0.5in}\left.  -60R\left(  G\right)  \mathcal{E}%
+180\mathcal{E}^{2}+30\mathbf{\Omega}_{\mu\nu}\mathbf{\Omega}_{\rho\sigma
}G^{\mu\rho}G^{\nu\sigma}\right]  ,
\end{align*}
where we have omitted the total derivative terms $12\left(  -R\left(
G\right)  +5\mathcal{E}\right)  _{;\mu}^{\quad\mu}$. Since the modification
from $\overline{\omega}_{\mu}$ to $\overline{\omega^{\prime}}_{\mu}$ is
abelian \eqref{fundrel} we get%
\[
\mathbf{\Omega}_{\mu\nu}=\Omega_{\mu\nu}%
\]
and
\[
\mathbf{\Omega}_{\mu\nu}\mathbf{\Omega}_{\rho\sigma}G^{\mu\rho}G^{\nu\sigma
}=e^{-4\phi}\Omega_{\mu\nu}\Omega_{\rho\sigma}g^{\mu\rho}g^{\nu\sigma}.
\]
Next we group the terms
\[
180\,\mathcal{E}^{2}-60\mathcal{E}R(G)+5R^{2}\left(  G\right)  =180\left(
\mathcal{E}-\frac{1}{6}R\left(  G\right)  \right)  ^{2},
\]
which yields upon using equation( \ref{invscale})
\[
180\;e^{-4\phi}\left(  E-\frac{1}{6}R\left(  g\right)  \right)  ^{2}.
\]
We are left with the terms
\[
30\;\mathbf{\Omega}_{\mu\nu}\mathbf{\Omega}_{\rho\sigma}G^{\mu\rho}%
G^{\nu\sigma}-2R_{\mu\nu}(G)R^{\mu\nu}\left(  G\right)  +2R_{\mu\nu\rho\sigma
}\left(  G\right)  R^{\mu\nu\rho\sigma}(G).
\]
We now use
\[
30\,Tr\left(  \mathbf{\Omega}_{\mu\nu}\mathbf{\Omega}_{\rho\sigma}G^{\mu\rho
}G^{\nu\sigma}\right)  =30\,e^{-4\phi}Tr\left(  \Omega_{\mu\nu}\Omega
_{\rho\sigma}g^{\mu\rho}g^{\nu\sigma}\right)  .
\]%
\[
-2R_{\mu\nu}(G)R^{\mu\nu}\left(  G\right)  +2R_{\mu\nu\rho\sigma}\left(
G\right)  R^{\mu\nu\rho\sigma}(G)=-R(G)^{\ast}R^{\ast}(G)+3\,C_{\mu\nu
\rho\sigma}\left(  G\right)  C^{\mu\nu\rho\sigma}\left(  G\right)  .
\]
In deriving the last relation we made use of the two identities
\begin{align*}
R^{\ast}R^{\ast}-R^{2}  &  =R_{\mu\nu\rho\sigma}^{2}-4R_{\mu\nu}^{2},\\
C_{\mu\nu\rho\sigma}^{2}-\frac{1}{3}R^{2}  &  =R_{\mu\nu\rho\sigma}%
^{2}-2R_{\mu\nu}^{2},
\end{align*}
where $C_{\mu\nu\rho\sigma}$ is the conformal tensor and
\[
R^{\ast}R^{\ast}=\frac{1}{4\sqrt{g}}\epsilon^{\mu\nu\rho\sigma}\epsilon
_{\alpha\beta\gamma\delta}R_{\mu\nu}^{\quad\alpha\beta}R_{\rho\sigma}%
^{\quad\gamma\delta}.
\]
These imply
\begin{align*}
R_{\mu\nu\rho\sigma}^{2}  &  =2C_{\mu\nu\rho\sigma}^{2}-R^{\ast}R^{\ast}%
+\frac{1}{3}R^{2},\\
R_{\mu\nu}^{2}  &  =\frac{1}{2}C_{\mu\nu\rho\sigma}^{2}-\frac{1}{2}R^{\ast
}R^{\ast}+\frac{1}{3}R^{2}.
\end{align*}
The square of the conformal tensor is known to be conformal invariant
\[
\int d^{4}x\sqrt{G}\,C_{\mu\nu\rho\sigma}\left(  G\right)  C^{\mu\nu\rho
\sigma}\left(  G\right)  =\int d^{4}x\sqrt{g}\,C_{\mu\nu\rho\sigma}\left(
g\right)  C^{\mu\nu\rho\sigma}\left(  g\right)  .
\]
The topological Gauss-Bonnet term is metric independent and therefore
conformal invariant
\[
\int d^{4}x\sqrt{G}\,R\left(  G\right)  ^{\ast}R^{\ast}\left(  G\right)  =\int
d^{4}x\sqrt{g}\,R\left(  g\right)  ^{\ast}R^{\ast}\left(  g\right)  ,
\]
and this can be rewritten as
\[
\frac{1}{4}\int d^{4}x\frac{1}{\sqrt{g}}\epsilon^{\mu\nu\rho\sigma}%
\epsilon_{\alpha\beta\gamma\delta}R_{\mu\nu}^{\quad\alpha\beta}R_{\rho\sigma
}^{\quad\gamma\delta}.
\]
This shows that the $a_{4}$ term has the expected invariance under the
rescaling of the operator $P_{0}\rightarrow P=\,e^{-\phi}\,P_{0}\,e^{-\phi}$.

\section{\bigskip Spectral action with dilaton}

We now use the result of the previous section to compute the spectral action
with dilaton as a function of the rescaled metric $G$ in the Einstein frame,
where the dilaton factor is absorbed in the metric.

The lowest term in the spectral action is given by:
\[
\frac{45}{4\pi^{2}}f_{0}\,\int d^{4}x\sqrt{g}e^{4\phi}=\,\frac{45}{4\pi^{2}%
}f_{0}\int d^{4}x\sqrt{G}.
\]
(Note that the dimension of the bundle on which the operator is acting is
$4\times3\times15$ where the $4$ is the dimension of spinors, $3$ the number
of generations and $15= 4\times3 +3$ is the content of each generation).

The next term in the spectral action with dilaton of the standard model is, in
terms of the original metric $g$ :
\begin{equation}
\frac{3}{4\pi^{2}}f_{2}\int d^{4}x\sqrt{g}e^{2\phi}\left(  \frac{5}{4}R\left(
g\right)  -2y^{2}H^{\ast}H\right)  . \label{a2term0}%
\end{equation}
We can transform this back to the Einstein frame with metric $G_{\mu\nu}$ so
that the curvature scalar term has no scale factors in front of it. \ Using
\eqref{scalar} with $g\rightarrow G$ and $\phi\rightarrow-\phi$ the curvature
$R\left(  g\right)  $ is
\[
R\left(  g\right)  =e^{2\phi}\left(  R\left(  G\right)  +6\,G^{\mu\nu}\left(
-\nabla_{\mu}^{G}\nabla_{\nu}^{G}\phi+\partial_{\mu}\phi\partial_{\nu}%
\phi\right)  \right)  ,
\]
and we obtain
\begin{align*}
\int d^{4}x\sqrt{g}e^{2\phi}R\left(  g\right)   &  =\int d^{4}x\sqrt{G}\left(
R\left(  G\right)  +6\,G^{\mu\nu}\left(  -\nabla_{\mu}^{G}\nabla_{\nu}^{G}%
\phi+\partial_{\mu}\phi\partial_{\nu}\phi\right)  \right) \\
&  =\int d^{4}x\sqrt{G}\left(  R\left(  G\right)  +6\,G^{\mu\nu}\partial_{\mu
}\phi\partial_{\nu}\phi\right)
\end{align*}
after integrating by parts. The $a_{2}$ term \eqref{a2term0} thus becomes
\begin{equation}
\frac{3}{4\pi^{2}}f_{2}\int d^{4}x\sqrt{G}\left(  \frac{5}{4}R\left(
G\right)  +\frac{15}{2}G^{\mu\nu}\partial_{\mu}\phi\,\partial_{\nu}\phi
-2y^{2}H^{^{\prime}\ast}H^{^{\prime}}\right)  \label{a2term}%
\end{equation}
where we have defined
\[
H=e^{\phi}H^{\prime},
\]
so that the only appearance of the dilaton $\phi$ is through its kinetic energy.

Let us pause a bit and discuss signs at this point. For a positive test
function $F$ the coefficients $f_{0}$, $f_{2}$, $f_{4}$ are all positive. It
is important that the Einstein term $\int d^{4}x\sqrt{G}\,R\left(  G\right)  $
appears in \eqref{a2term} with the correct sign for the Euclidean functional
integral, and that the kinetic term for $\phi$ namely $\int d^{4}x\sqrt
{G}\,G^{\mu\nu}\partial_{\mu}\phi\partial_{\nu}\phi$ appears with a positive
coefficient in \eqref{a2term}.

The next term coming from $a_{4}\left(  x,P\right)  $ is unchanged for the
spectral action with dilaton, and thus given independently of $\phi$ by
\cite{ACAC},
\begin{align*}
&  \frac{f_{4}}{4\pi^{2}}\int d^{4}x\sqrt{g}\left(  \frac{1}{32}\left(
11R\left(  g\right)  \,^{\ast}R^{\ast}\left(  g\right)  -18C_{\mu\nu\rho
\sigma}\left(  g\right)  C_{\alpha\beta\gamma\delta}\left(  g\right)
g^{\mu\alpha}g^{\nu\beta}g^{\rho\gamma}g^{\sigma\delta}\right)  \right. \\
&  \hspace{1in}+3y^{2}\left(  D_{\mu}H^{\ast}D_{\nu}Hg^{\mu\nu}-\frac{1}%
{6}R\left(  g\right)  H^{\ast}H\right) \\
&  \qquad\qquad\qquad\left.  +\left(  g_{3}^{2}G_{\mu\nu}^{i}G_{\rho\sigma
}^{i}+g_{2}^{2}F_{\mu\nu}^{\alpha}F_{\rho\sigma}^{\alpha}+\frac{5}{3}g_{1}%
^{2}B_{\mu\nu}B_{\rho\sigma}\right)  g^{\mu\rho}g^{\nu\sigma}+3z^{2}\left(
H^{\ast}H\right)  ^{2}\right)  ,
\end{align*}
where we\ have omitted total derivatives as they only contribute to boundary
terms. Let us show that we can rewrite this term in the following way as a
function of the metric $G_{\mu\nu}$ by making use of the conformal invariance
of $a_{4}$:
\begin{align*}
&  \frac{f_{4}}{4\pi^{2}}\int d^{4}x\sqrt{G}\left(  \frac{1}{32}\left(
11R\left(  G\right)  \,^{\ast}R^{\ast}\left(  G\right)  -18C_{\mu\nu\rho
\sigma}\left(  G\right)  C_{\alpha\beta\gamma\delta}\left(  G\right)
G^{\mu\alpha}G^{\nu\beta}G^{\rho\gamma}G^{\sigma\delta}\right)  \right. \\
&  \hspace{1in}+3y^{2}\left(  D_{\mu}H^{^{\prime}\ast}D_{\nu}H^{\prime}%
G^{\mu\nu}-\frac{1}{6}R\left(  G\right)  H^{\prime\ast}H^{\prime}\right) \\
&  \qquad\qquad\qquad\left.  +\left(  g_{3}^{2}G_{\mu\nu}^{i}G_{\rho\sigma
}^{i}+g_{2}^{2}F_{\mu\nu}^{\alpha}F_{\rho\sigma}^{\alpha}+\frac{5}{3}g_{1}%
^{2}B_{\mu\nu}B_{\rho\sigma}\right)  G^{\mu\rho}G^{\nu\sigma}+3z^{2}\left(
H^{^{\prime}\ast}H^{\prime}\right)  ^{2}\right)  .
\end{align*}

The terms which only involve the metric are conformal by construction. The
same holds for the terms which involve the gauge fields since the Yang-Mills
action is conformal. Thus we need only to take care of the terms that involve
the Higgs fields. We have to show that the following expression is unchanged
by $g\rightarrow G$ and $H\rightarrow H^{\prime}$,
\[
\frac{3\,f_{4}\,y^{2}}{4\pi^{2}}\,\int d^{4}x\sqrt{g}\left(  g^{\mu\nu}D_{\mu
}H^{\ast}D_{\nu}H-\frac{1}{6}R\left(  g\right)  H^{\ast}H\right)  .
\]
To see this we first rescale the kinetic energy of the Higgs field
\begin{align*}
\sqrt{g}\,g^{\mu\nu}D_{\mu}H^{\ast}D_{\nu}H  &  =\sqrt{G}G^{\mu\nu}e^{-2\phi
}D_{\mu}\left(  e^{\phi}H^{\prime\ast}\right)  D_{\nu}\left(  e^{\phi
}H^{\prime}\right) \\
&  =\sqrt{G}G^{\mu\nu}\left(  D_{\mu}H^{\prime\ast}D_{\nu}H^{^{\prime}%
}+\partial_{\mu}\phi H^{\prime\ast}D_{\nu}H^{\prime}\right. \\
&  \left.  +D_{\mu}H^{\prime\ast}H^{\prime}\partial_{\nu}\phi+H^{\prime\ast
}H^{\prime}\partial_{\mu}\phi\partial_{\nu}\phi\right)  .
\end{align*}
The conformal coupling of the Higgs field to the scalar curvature transforms
as%
\[
-\frac{1}{6}\sqrt{g}\,R(g)H^{\ast}H=-\frac{1}{6}\sqrt{G}\,H^{\prime\ast
}H^{\prime}\left(  R(G)-6G^{\mu\nu}\left(  \left(  \nabla_{\mu}\nabla_{\nu
}\right)  ^{G}\phi-\partial_{\mu}\phi\partial_{\nu}\phi\right)  \right)  .
\]
After integrating by parts the term
\[
\int d^{4}x\sqrt{G}\,H^{\prime\ast}H^{\prime}\left(  \nabla_{\mu}\nabla_{\nu
}\right)  ^{G}\phi\,G^{\mu\nu},
\]
we find that all cross terms cancel, thus obtaining
\begin{align*}
&  \int d^{4}x\sqrt{g}\left(  g^{\mu\nu}D_{\mu}H^{\ast}D_{\nu}H-\frac{1}%
{6}R\left(  g\right)  H^{\ast}H\right) \\
&  =\int d^{4}x\sqrt{G}\left(  G^{\mu\nu}D_{\mu}H^{\prime\ast}D_{\nu}%
H^{\prime}-\frac{1}{6}R\left(  G\right)  H^{\prime\ast}H^{\prime}\right)  .
\end{align*}
The quartic Higgs interactions are evidently scale invariant
\[
\int d^{4}x\sqrt{g}\left(  H^{\ast}H\right)  ^{2}=\int d^{4}x\sqrt{G}\left(
H^{\prime\ast}H^{\prime}\right)  ^{2}.
\]
Collecting all terms, the low-energy bosonic part of the spectral action with
dilaton is given by%
\begin{align}
I_{b}  &  =\frac{45}{4\pi^{2}}f_{0}\int d^{4}x\sqrt{G}\\
&  +\frac{3}{4\pi^{2}}f_{2}\int d^{4}x\sqrt{G}\left(  \frac{5}{4}R\left(
G\right)  +\frac{15}{2}G^{\mu\nu}\partial_{\mu}\phi\partial_{\nu}\phi
-2y^{2}H^{^{\prime}\ast}H^{^{\prime}}\right) \nonumber\\
&  +\frac{f_{4}}{4\pi^{2}}\int d^{4}x\sqrt{G}\left(  \frac{1}{32}\left(
11R\left(  G\right)  \,^{\ast}R^{\ast}\left(  G\right)  -18\,C_{\mu\nu
\rho\sigma}\left(  G\right)  C^{\mu\nu\rho\sigma}\left(  G\right)  \right)
\right. \nonumber\\
&  \hspace{1in}+3y^{2}\left(  D_{\mu}H^{^{\prime}\ast}D_{\nu}H^{\prime}%
G^{\mu\nu}-\frac{1}{6}R\left(  G\right)  H^{\prime\ast}H^{\prime}\right)
\nonumber\\
&  \qquad\left.  +\left(  g_{3}^{2}G_{\mu\nu}^{i}G_{\rho\sigma}^{i}+g_{2}%
^{2}F_{\mu\nu}^{\alpha}F_{\rho\sigma}^{\alpha}+\frac{5}{3}g_{1}^{2}B_{\mu\nu
}B_{\rho\sigma}\right)  G^{\mu\rho}G^{\nu\sigma}+3z^{2}\left(  H^{^{\prime
}\ast}H^{\prime}\right)  ^{2}\right)  .\nonumber
\end{align}

For higher order terms one expects a scaling factor of the form $e^{\left(
4-n\right)  \phi}$ to be present, but derivatives of the dilaton field $\phi$
may also occur. Therefore in the Einstein frame, one does no expect the
dilaton field $\phi$ to acquire a potential. As will be discussed later, this
will change when quantum corrections are taken into account and the dilaton
acquires a potential of the Coleman-Weinberg type \cite{CW}.

Fermionic interactions take the simple form
\[
\left\langle \Psi\left\vert D\right\vert \Psi\right\rangle =\int d^{4}%
x\sqrt{g}\,\overline{\Psi}D\Psi,
\]
where the metric $g_{\mu\nu}$ is used to insure hermiticity of $D$. We will
now show that the fermions will not feel the dilaton. To see this we first
redefine the spinors by
\[
\Psi=e^{\frac{3}{2}\phi}\Psi^{\prime},
\]
then we have, for the parts not involving the Higgs or gauge fields,
\[
\left\langle \Psi\left\vert D\right\vert \Psi\right\rangle =\int d^{4}%
x\sqrt{G}e^{-4\phi}e^{\frac{3}{2}\phi}\overline{\Psi}^{\prime}\gamma
^{c}e^{\phi}E_{c}^{\mu}\left(  \partial_{\mu}+\frac{1}{4}\omega_{\mu}%
^{\,ab}\left(  e\right)  \gamma_{ab}\right)  \left(  e^{\frac{3}{2}\phi}%
\Psi^{\prime}\right)  ,
\]
where the rescaled vierbein is $E_{\mu}^{a}=e^{\phi}e_{\mu}^{a}$ . \ We have
to express the spin-connection $\omega_{\mu}^{\,ab}\left(  e\right)  $ in
terms of the spin-connection of the rescaled vierbein $\Omega_{\mu}%
^{ab}\left(  E\right)  .$ To do this we use the equations
\begin{align*}
\partial_{\mu}e_{\nu}^{a}-\partial_{\nu}e_{\mu}^{a}-\omega_{\mu}^{\,ab}\left(
e\right)  e_{\nu b}+\omega_{\nu}^{\,ab}\left(  e\right)  e_{\mu b}  &  =0,\\
\partial_{\mu}E_{\nu}^{a}-\partial_{\nu}E_{\mu}^{a}-\Omega_{\mu}^{\,ab}\left(
E\right)  E_{\nu b}+\Omega_{\nu}^{\,ab}\left(  E\right)  E_{\mu b}  &  =0,
\end{align*}
and these imply that
\[
\omega_{\mu}^{\,ab}\left(  e\right)  =\Omega_{\mu}^{\,ab}\left(  E\right)
+\left(  e_{\mu}^{a}e^{\nu b}-e_{\mu}^{a}e^{\nu b}\right)  \partial_{\nu}%
\phi.
\]
Therefore
\[
\gamma^{c}e_{c}^{\mu}\left(  \frac{3}{2}\partial_{\mu}\phi+\frac{1}{4}%
\omega_{\mu}^{\,ab}\left(  e\right)  \gamma_{ab}\right)  =\gamma^{c}e^{\phi
}E_{c}^{\mu}\left(  \frac{1}{4}\Omega_{\mu}^{\,ab}\left(  E\right)
\gamma_{ab}\right)  ,
\]
and the fermionic action reduces to the nice form
\[
\int d^{4}x\sqrt{G}\,\overline{\Psi}^{\prime}\gamma^{c}E_{c}^{\mu}\left(
\partial_{\mu}+\frac{1}{4}\Omega_{\mu}^{\,ab}\left(  E\right)  \gamma
_{ab}\right)  \Psi^{\prime}%
\]
which is independent of the dilaton. Finally the parts involving interactions
between the fermions and the Higgs or gauge fields could be written in the
form
\begin{align*}%
{\displaystyle\int}
d^{4}x\sqrt{g}\,\overline{\Psi}\gamma_{5}H\Psi &  =%
{\displaystyle\int}
d^{4}x\sqrt{G}\,\overline{\Psi^{\prime}}\gamma_{5}H^{\prime}\Psi^{^{\prime}%
},\\%
{\displaystyle\int}
d^{4}x\sqrt{g}\,\overline{\Psi}\gamma^{a}e_{a}^{\mu}A_{\mu}\Psi &  =%
{\displaystyle\int}
d^{4}x\sqrt{G}\,\overline{\Psi^{\prime}}\gamma^{a}E_{a}^{\mu}A_{\mu}%
\Psi^{^{\prime}}.
\end{align*}

The fermionic interactions are
\[
I_{f}=\left\langle Q\left\vert D_{q}\right\vert Q\right\rangle +\left\langle
L\left\vert D_{l}\right\vert L\right\rangle
\]
where
\[
Q=\left(
\begin{array}
[c]{c}%
u_{L}\\
d_{L}\\
d_{R}\\
u_{R}%
\end{array}
\right)  ,\quad\left(
\begin{array}
[c]{c}%
\nu_{L}\\
e_{L}\\
e_{R}%
\end{array}
\right)
\]
and these take exactly the same form as those without dilaton when expressed
in terms of the metric $G_{\mu\nu}$. Rewriting this in terms of the fermionic
fields
\[
Q^{\prime}=e^{-\frac{3}{2}\phi}Q,\qquad L^{\prime}=e^{-\frac{3}{2}\phi}L,
\]
and the Higgs field $H^{\prime}$ we obtain%
\[
I_{f}=\int d^{4}x\sqrt{G}\left(  \overline{L}^{^{\prime}}D_{l}^{^{\prime}%
}L^{\prime}+\overline{Q}^{\prime}D_{q}^{^{\prime}}Q^{^{\prime}}\right)
\]
where \cite{ACAC}
\[
D_{l}^{^{\prime}}=\left(
\begin{array}
[c]{cc}%
\gamma^{\mu}\otimes\left(  D_{\mu}\otimes1_{2}-\frac{i}{2}g_{2}A_{\mu}%
^{\alpha}\sigma^{\alpha}+\frac{i}{2}g_{1}B_{\mu}\otimes1_{2}\right)
\otimes1_{3} & \gamma_{5}\otimes k^{e}\otimes H^{^{\prime}}\\
\gamma_{5}\otimes k^{\ast e}\otimes H^{\ast^{\prime}} & \gamma^{\mu}%
\otimes\left(  D_{\mu}+ig_{1}B_{\mu}\right)  \otimes1_{3}%
\end{array}
\right)
\]%
\begin{align*}
D_{q}^{\prime}  &  =\left(
\begin{array}
[c]{ccc}%
\gamma^{\mu}\otimes\nabla_{\mu}^{\left(  1,2\right)  }\otimes1_{3} &
\gamma_{5}\otimes k^{d}\otimes H^{^{\prime}} & \gamma_{5}\otimes k^{u}%
\otimes\widetilde{H}^{\prime}\\
\gamma_{5}\otimes k^{\ast d}\otimes H^{\ast^{\prime}} & \gamma^{\mu}%
\otimes\left(  D_{\mu}+\frac{i}{3}g_{1}B_{\mu}\right)  \otimes1_{3} & 0\\
\gamma_{5}\otimes k^{\ast u}\otimes\widetilde{H}^{\prime\ast} & 0 &
\gamma^{\mu}\otimes\left(  D_{\mu}-\frac{2i}{3}g_{1}B_{\mu}\right)
\otimes1_{3}%
\end{array}
\right) \\
&  +\gamma^{\mu}\otimes1_{4}\otimes1_{3}\otimes\left(  -\frac{i}{2}g_{3}%
V_{\mu}^{i}\lambda^{i}\right)  ,
\end{align*}
and
\begin{align*}
\gamma^{\mu}  &  =\gamma^{a}E_{a}^{\mu},\\
D_{\mu}  &  =\partial_{\mu}+\frac{1}{4}\Omega_{\mu}^{\,ab}\left(  E\right)
\gamma_{ab},\\
\nabla_{\mu}^{\left(  1,2\right)  }  &  =D_{\mu}\otimes1_{2}-\frac{i}{2}%
g_{2}A_{\mu}^{\alpha}\sigma^{\alpha}-\frac{i}{6}g_{1}B_{\mu}\otimes1_{2}.
\end{align*}
From the above considerations we deduce that the only effect of the dilaton on
the low-energy terms of the spectral action is that the dilaton gets a kinetic
term with no other interactions. \ This confirms that all matter interactions
in the above Lagrangian are scale invariant when expressed in the rescaled
fields $G_{\mu\nu}$, $H^{\prime}$ and $\Psi^{\prime}$. Only the Einstein term
and the dilaton kinetic energy are not scale invariant.

Note that the invariance of the action for the Fermions, that is the equality
\begin{equation}
\left\langle \Psi\left\vert D\right\vert \Psi\right\rangle =\,\left\langle
\Psi^{\prime}\left\vert D^{\prime}\right\vert \Psi^{\prime}\right\rangle
^{\prime}%
\end{equation}
where $D^{\prime}$ corresponds to the metric $G$ and the fields $H^{\prime}$,
does not mean that the operators $D$ and $D^{\prime}$ are the same. Indeed the
transformation $\Psi\rightarrow\Psi^{\prime}$ is not unitary and one has
\begin{equation}
\left\langle \Psi^{\prime}|\Psi^{\prime}\right\rangle ^{\prime}=\,\left\langle
\Psi\left\vert e^{\phi}\right\vert \Psi\right\rangle
\end{equation}
which gives the unitary equivalence
\begin{equation}
D^{\prime}\sim\,e^{-\phi/2}\,D\,e^{-\phi/2}%
\end{equation}
One might then be tempted to conclude that the square of $e^{-\phi
/2}\,D\,e^{-\phi/2}$ should be unitarily equivalent to $P=\,e^{-\phi}%
\,D^{2}\,e^{-\phi}$ but this does not hold precisely because of the additional
kinetic term in the spectral action with dilaton. Indeed one can prove
(Appendix C) in the general framework of spectral triples, with a minimum
amount of hypothesis, the identity
\begin{equation}
\label{kin}\int\!\!\!\!\!\!-\ \,e^{2\phi}\,D^{-2}=\,\int
\!\!\!\!\!\!-\ \,(e^{-\phi/2}\,D\,e^{-\phi/2})^{-2}\,+\,\frac{1}{2}%
\int\!\!\!\!\!\!-\ \,[D,\,e^{\phi}][D,\,e^{\phi}]^{\ast}\,D^{-4}%
\end{equation}
with $D^{\prime}=e^{-\phi/2}\,D\,e^{-\phi/2}$ the last term gives the
canonical kinetic energy of the dilaton
\[
\frac{1}{2}\int\!\!\!\!\!\!-\ \,[D^{\prime},\,\phi][D^{\prime},\,\phi]^{\ast
}\,D^{^{\prime}\,-4}
\]
with the correct sign.

\section{\bigskip Applications}

We have shown that the dilaton interactions of the spectral action are almost
the same as the ones proposed in the literature \cite{Bud}, \cite{Bubu}, the
difference lies in the derivative couplings of the dilaton field. These were
obtained by requiring the standard model matter sector to be scale invariant
by introducing a compensating dilaton field. The origin of the dilatational
symmetry breaking are the mass terms of the Higgs potential, and these are
scaled with the dilaton field to make them scale invariant. In a curved
space-time all fields couple to gravity, and the dilaton. The proposed action
for the gravity-dilaton-Higgs sectors, in our notation, was derived to
be\footnote{This expression is in the conventions of \cite{Kolb} and is in
Minkowski space.} \cite{Bubu}, \cite{Kolb}
\begin{align*}
I  &  =%
{\displaystyle\int}
d^{4}x\sqrt{G}\left(  -\frac{1}{2\kappa^{2}}R+\frac{1}{2}\left(  1+\frac
{6}{\kappa^{2}f^{2}}\right)  G^{\mu\nu}\partial_{\mu}\phi\partial_{\nu}%
\phi\right. \\
&  \qquad\qquad\qquad\left.  +G^{\mu\nu}D_{\mu}H^{^{\prime}\ast}D_{\nu
}H^{^{\prime}}-V_{0}\left(  H^{^{\prime}\ast}H^{^{\prime}}\right)  .\right)
\end{align*}
There it was shown that in curved space-time this corresponds to the
Jordan-Brans-Dicke theory of gravity. The only difference between this action
and the spectral action is that the later has the conformal coupling
\[
\frac{1}{6}R(G)\left(  H^{^{\prime}\ast}H^{^{\prime}}\right)  ^{2},
\]
which is necessary to make the matter couplings scale invariant. A slight
modification was also proposed in the study of models of extended inflation,
by also considering the possibility of modifying the Higgs sector by taking
\cite{Kolb}%
\[
e^{\frac{2}{f}\phi}G^{\mu\nu}D_{\mu}H^{\ast}D_{\nu}H-e^{\frac{4}{f}\phi}%
V_{0}\left(  H^{\ast}H\right)
\]
This differs from the spectral action by the appearance of derivative
couplings of the form
\[
G^{\mu\nu}D_{\mu}H^{^{\prime}\ast}H^{^{\prime}}\partial_{\nu}\phi.
\]
It is amusing to note that this alternative proposed action is exactly the
same action as the one derived for the Connes-Lott gravitational interactions
\cite{CFF}, \cite{Lizzi}. Therefore the two models proposed in the literature
for making the Higgs sector scale invariant are the same as the interactions
obtained for the noncommutative standard model, either for the spectral action
formulation, or the Connes-Lott formulation. We also note that scale
invariance of the action is broken by the Einstein term and by the kinetic
term for the dilaton. This is remarkable because it was shown that if the full
action is scale invariant, then the couplings will not lead to a model with
extended inflation. Quantum corrections and renormalization conditions break
scale invariance in the matter sector of the standard model and lead to an
exponentially large hierarchy between the mass scale $f$ \ where $\phi
=\frac{1}{f}\sigma$ and the electroweak scale without fine tuning. The scale
$f$ \ is normally of the order of the Planck scale. The dilaton mass obtained
depends on the Higgs mass, but should be constrained to be smaller than
$10^{-6}$ev. \ 

The noncommutative space of the standard model is obtained by taking the
product of a four-dimensional Riemannian manifold times a discrete space
dictated by the symmetries of the Hilbert space spanned by the quarks and
leptons. The presence of left and right handed fermions provides the intuitive
picture where these fermions are placed on two different sheets. The gauge
fields in the discrete dimensions are the Higgs fields, with the inverse of
the distance between the two sheets interpreted as the electroweak energy
scale. This picture is similar to the Randall-Sundrum (RS) scenario where the
four-dimensional space is embedded into a five-dimensional space as a 3-brane
positioned at the points $x_{5}=0$ and $x_{5}=\pi r_{c}$ where $r_{c}$ is the
compactification radius. The action for the Higgs sector in the RS model
\cite{RS} was obtained to be
\begin{align*}
&
{\displaystyle\int}
d^{4}x\sqrt{g}\left(  g^{\mu\nu}D_{\mu}H^{\ast}D_{\nu}H-\lambda\left(
\left\vert H\right\vert ^{2}-v_{0}^{2}\right)  \right) \\
=  &
{\displaystyle\int}
d^{4}x\sqrt{\overline{g}}\left(  \overline{g}^{\mu\nu}D_{\mu}H^{^{\prime}\ast
}D_{\nu}H^{^{\prime}}-\lambda\left(  \left\vert H^{^{\prime}}\right\vert
^{2}-e^{-2kr_{c}\pi}v_{0}^{2}\right)  \right)
\end{align*}
where
\begin{align*}
g^{\mu\nu}  &  =e^{-2kr_{c}\pi}\overline{g}^{\mu\nu},\\
H^{^{\prime}}  &  =e^{2kr_{c}\phi}H,
\end{align*}
in the visible sector located at $x_{5}=\pi r_{c}.$ The physical mass scales
are set by the symmetry breaking scale $v=e^{-kr_{c}\pi}v_{0}$ so that
$m=m_{0}e^{-kr_{c}\pi}.$ The bare symmetry breaking scale $v_{0}$ is taken to
be of the order of the Planck scale at $10^{19}$ Gev and the scaling factor
$e^{kr_{c}\pi}$ tuned to be of the order of $10^{15}$ so that the low-energy
masses are of the order of Tev. The hierarchy problem is only partially solved
in a technical sense because the tuning could not be maintained at the quantum
level. A choice of $kr_{c}=10$ can generate the large scale $10^{15}$ Gev.
Comparing the Higgs sectors in the RS action with that in the spectral action
we immediately see that \ they are identical provided we identify the
expectation value of the dilaton field $\left\langle \phi\right\rangle $ with
$kr_{c}\pi.$

\section{Conclusions}

The Dirac operator being a differential operator has the dimensions of mass.
The spectral action in noncommutative geometry is defined as a function of a
dimensionless operator which is taken to be the Dirac operator divided by some
arbitrary large mass scale. The arbitrariness of the mass scale naturally
suggests to make this scale dynamical by introducing a dilaton field in the
Dirac operator of the noncommutative space defined by the standard model. To
understand the appearance of the mass scales of the spectral action, we
evaluated all interactions of the dilaton with the matter sector in the
standard model. We found the remarkable result that the low-energy action,
when evaluated in the Einstein frame, is scale invariant except for the
Einstein-Hilbert term and the dilaton kinetic term. The resulting model is
almost identical to the one proposed in the literature \cite{Bubu}%
,\cite{Bud},\cite{Kolb}. The main motivation in these works is the observation
that the standard model is classically almost scale invariant, with the
symmetry only broken by the mass term in the Higgs potential. The symmetry is
restored by the use of a dilaton field. When coupled to gravity, neither the
dilaton kinetic energy nor the scalar curvature are scale invariant, leading
to a Jordan-Brans-Dicke theory of gravity. The vacuum expectation value of the
Higgs field is then dependent on the dilaton and is classically undetermined.
Quantum corrections break the scale invariance of the scalar potential and
change the vacuum expectation value of the Higgs field. \ The dilaton acquires
a large negative expectation value given by $-m$ and a small mass. The
hierarchy in mass scales is due to the large Yukawa coupling of the top quark.
The dilaton expectation value can range between the GUT scale of $10^{15}$ Gev
to the Planck scale of $2.4\cdot10^{18}$ Gev. \ The hierarchy in mass scales
is not possible if the dilaton kinetic energy and the gravitational action
were scale invariant. It is remarkable that all the essential features of
building a scale invariant standard model interactions to generate a mass
hierarchy and predict the Higgs mass are naturally included in the spectral
action without any fine tuning. It is worth mentioning that the scalar
potential of exactly the same model considered here was shown to admit
extended inflation and a metastable ground state. It also evades the problems
of the original version of extended inflation.

The vacuum expectation value of the dilaton field is determined by getting
contributions from classical and radiative corrections to the vacuum energy
density. \ One does not obtain naturally a vanishing cosmological constant.
There are two possibilities to cure this problem. The first is to determine
the low-energy value of the cosmological constant as determined by the
renormaliztion group equations and then fine tune this value to cancel the
contributions of the Coleman-Weinberg potential. The second possibility to
cure this problem is to fix the total invariant volume. This restricts general
relativity to the form considered in \cite{Bud2} \ where the volume form is
held fixed. There it was shown that this picture is consistent both at the
classical and quantum levels \cite{van}, \cite{dragon},\cite{Ng} . This fixes
the total invariant volume and eliminates the scalar mode of the metric tensor
$g_{\mu\nu}.$ This is done at the expense of introducing the dilaton mode
$\phi$ \cite{Bubu}. In noncommutative geometry the volume is fixed by a
Hochschild cycle $c$ whose compatibility with the Dirac operator $D$ is a
basic constraint on the Hilbert space representation giving the metric
\cite{Alain}. One applies the representation to monomials
\[
\pi\left(  f_{0},f_{1},f_{2},f_{3},f_{4}\right)  =\,f_{0}\left[
D,f_{1}\right]  \left[  D,f_{2}\right]  \left[  D,f_{3}\right]  \left[
D,f_{4}\right]  ,
\]
and requires that when applied to the Hochschild cycle $c$ it gives
\[
\pi\left(  c\right)  =\gamma_{5}.
\]
The cosmological constant becomes determined by the initial conditions of the theory.

To summarize, we have shown that the spectral action includes naturally a
dilaton field which guarantees the scale invariance of the standard model
interactions, and provides a mechanism to generate mass hierarchies. This is
in addition to the advantages obtained previously in \cite{ACAC} \ which are
now well known \cite{Kastler}. \ There it was shown that all the correct
features of the standard model are obtained without any fine tuning, such as
unification with gravity, unification of the three gauge coupling constants
and relating the Higgs to the gauge couplings. These results should be taken
to support the idea that all the geometric information about the physical
space is captured by the knowledge of the Dirac operator of an appropriate
noncommutative space.

\section{Appendix A}

In this appendix we shall prove formula \eqref{commutator}. Given an elliptic
positive invertible second order operator $Q$ and a differential operator $T$
we use the notation $\nabla(T)=\,[Q,T]$ and the following identity (for $n
\geq0$) :
\begin{equation}
\label{pass}Q^{-1}\,T=\,\sum_{0}^{n}\,(-1)^{k} \,\nabla^{k}(T)\,Q^{-k-1}%
\,+\,(-1)^{n+1}\,Q^{-1}\,\nabla^{n+1}(T)\,Q^{-n-1}\,.
\end{equation}
We apply this to $Q=\,P(t)+\lambda$, $T=\,\phi\,P(t)+\,P(t)\,\phi$. The
operator $\nabla^{k}(T)$ is independent of $\lambda$ and is a differential
operator of order $\leq2 + k$ since $\nabla^{k}(\phi)$ is at most of order
$k$. Thus the operator $\nabla^{k}(T)\,Q^{-k-1}$ is pseudodifferential of
order at most $2 + k -2(k+1)=-k$ and the remainder in \eqref{pass} is of order
at most $-n-1$. This shows that when working modulo operators of order less
than $-n$ we have
\[
(P(t)+\lambda)^{-1}\,T\sim\,T\,(P(t)+\lambda)^{-1}+\,\sum_{1}^{n}\,(-1)^{k}
\,\nabla^{k}(T)\,(P(t)+\lambda)^{-k-1}
\]
so that
\[
(P(t)+\lambda)^{-1}\,T\,(P(t)+\lambda)^{-1}\sim\,T\,(P(t)+\lambda)^{-2}
\]
\[
+\,\sum_{1}^{n}\,(-1)^{k} \,\nabla^{k}(T)\,(P(t)+\lambda)^{-k-2}\,.
\]
But one has
\[
\sum_{1}^{n}\,(-1)^{k} \,\nabla^{k}(T)\,(P(t)+\lambda)^{-k-2}%
=\,[P(t),A(t,\lambda)]
\]
where
\[
A(t,\lambda)=\,\sum_{1}^{n}\,(-1)^{k} \,\nabla^{k-1}(T)\,(P(t)+\lambda
)^{-k-2}
\]
Thus integrating from $\lambda=0$ to $\infty$ and using \eqref{derY} we get
that, modulo operators of order less than $-n$,
\[
\frac{d}{dt}\,Y(t)\sim\,\,-\,{\displaystyle\int\limits_{0}^{\infty}}\,
(\phi\,P(t)+\,P(t)\,\phi)\,(P(t)+\lambda)^{-2}\,d\lambda+\,[P(t),A(t)]
\]
where
\[
A(t)=\,-\,{\displaystyle\int\limits_{0}^{\infty}}\,A(t,\lambda)\,d\lambda\,.
\]
This thus gives
\[
\frac{d}{dt}\,Y(t)\sim\,\,-\, (\phi\,P(t)+\,P(t)\,\phi)\,P(t)^{-1}
+\,[P(t),A(t)]
\]
\[
=\, -2 \,\phi+\,[P(t),C(t)]\,,\quad C(t)=A(t)-\phi\,P(t)^{-1}\,.
\]
It is important to note that all of the above manipulations hold in the
general context of spectral triples with simple dimension spectrum. Moreover
one can prove fairly strong properties of the spectral action in this general context.

\section{Appendix B}

\bigskip In this appendix we shall prove the identity
\[
a_{n}\left(  x,P\right)  =a_{n}\left(  x,P_{1}\right)  =a_{n}\left(
x,P_{2}\right)
\]
where $P=e^{-\phi}D^{2}e^{-\phi}$ and $P_{1}=D^{2}e^{-2\phi},$ $P_{2}%
=e^{-2\phi}D^{2}$. It can then be used to simplify some of the computations of
section \ref{seeleydewitt}.

One simply writes
\[
P=e^{-\phi}P_{1}e^{\phi}%
\]
so that
\[
\mathrm{Trace}\left(  P^{-s}\right)  =\mathrm{Trace}\left(  P_{1}^{-s}\right)
\]
From this the identity $a_{n}\left(  x,P\right)  =a_{n}\left(  x,P_{1}\right)
$ immediately follows.

One can also do a direct check as follows, one first writes
\begin{align*}
P  &  =-\left(  G^{\mu\nu}I\,\partial_{\mu}\partial_{\nu}+\mathcal{A}^{\mu
}\partial_{\mu}+\mathcal{B}\right) \\
P_{1}  &  =-\left(  G^{\mu\nu}I\,\partial_{\mu}\partial_{\nu}+\mathcal{A}%
_{1}^{\mu}\partial_{\mu}+\mathcal{B}_{1}\right)  ,
\end{align*}
where%
\begin{align*}
G^{\mu\nu}  &  =e^{-2\phi}g^{\mu\nu}\\
\mathcal{B}  &  =e^{-2\phi}B+G^{\mu\nu}\left(  \partial_{\mu}\phi\partial
_{\nu}\phi-\partial_{\mu}\partial_{\nu}\phi\right)  -e^{-2\phi}A^{\mu}%
\partial_{\mu}\phi\\
\mathcal{A}^{\mu}  &  =e^{-2\phi}A^{\mu}-2G^{\mu\nu}\partial_{\nu}\phi,\\
\mathcal{A}_{1}^{\mu}  &  =e^{-2\phi}A^{\mu}-4G^{\mu\nu}\partial_{\nu}\phi\\
\mathcal{B}_{1}  &  =e^{-2\phi}B+2G^{\mu\nu}\left(  2\partial_{\mu}%
\phi\partial_{\nu}\phi-\partial_{\mu}\partial_{\nu}\phi\right)  -2e^{-2\phi
}A^{\mu}\partial_{\mu}\phi
\end{align*}
These relations imply
\begin{align*}
\mathcal{A}_{1}^{\mu}  &  =\mathcal{A}^{\mu}-2G^{\mu\nu}\partial_{\nu}\phi,\\
\mathcal{B}_{1}  &  =\mathcal{B+}G^{\mu\nu}\left(  \partial_{\mu}\phi
\partial_{\nu}\phi-\partial_{\mu}\partial_{\nu}\phi\right)  -\mathcal{A}^{\mu
}\partial_{\mu}\phi
\end{align*}
We also have
\begin{align*}
\overline{\omega^{\prime}}_{1\mu}  &  =\frac{1}{2}G_{\mu\nu}\left(
\mathcal{A}_{1}^{\nu}+\Gamma^{\nu}\left(  G\right)  \right) \\
&  =\overline{\omega^{\prime}}_{\mu}-\partial_{\mu}\phi
\end{align*}
so that
\begin{align*}
\mathcal{E}_{1}  &  =\mathcal{B}_{1}-G^{\mu\nu}\left(  \partial_{\mu}%
\overline{\omega^{\prime}}_{1\nu}+\overline{\omega^{\prime}}_{1\mu}%
\overline{\omega^{\prime}}_{1\nu}-\Gamma_{\mu\nu}^{\rho}\left(  G\right)
\overline{\omega^{\prime}}_{1\rho}\right) \\
&  =\mathcal{E}%
\end{align*}
Similarly
\[
\mathbf{\Omega}_{1\mu\nu}=\mathbf{\Omega}_{\mu\nu}%
\]
and the equality of the Seely-de Witt coefficients follow from the fact that
these depend only on $\mathcal{E},$ $\mathbf{\Omega}_{\mu\nu}$ and the
curvature tensors are functions of the same metric $G_{\mu\nu}.$

\section{Appendix C}

In this appendix we shall show \eqref{kin} and the appearance of the kinetic
term in the general framework of spectral triples using the following
manipulations. One has
\begin{equation}
\int\!\!\!\!\!\!-\ \,(e^{-\phi/2}\,D\,e^{-\phi/2})^{-2}=\,\int
\!\!\!\!\!\!-\ \,e^{\phi}\,D^{-1}\,e^{\phi}\,D^{-1}. \label{com0}%
\end{equation}
Also
\begin{equation}
D^{-1}\,e^{\phi}=\,e^{\phi}\,D^{-1}-\,D^{-1}\,[D,\,e^{\phi}]\,D^{-1},
\label{comd}%
\end{equation}
which allows to write \eqref{com0} as
\begin{equation}
\int\!\!\!\!\!\!-\ \,(e^{-\phi/2}\,D\,e^{-\phi/2})^{-2}=\,\int
\!\!\!\!\!\!-\ \,e^{2\phi}\,D^{-2}-\,\int\!\!\!\!\!\!-\ \,e^{\phi}%
\,D^{-1}\,[D,\,e^{\phi}]\,D^{-2}, \label{com1}%
\end{equation}
and using \eqref{comd} again,
\[
-\,\int\!\!\!\!\!\!-\ \,e^{\phi}\,D^{-1}\,[D,\,e^{\phi}]\,D^{-2}%
=\,-\,\int\!\!\!\!\!\!-\ \,e^{\phi}\,[D,\,e^{\phi}]\,D^{-3}-\,\int
\!\!\!\!\!\!-\ \,D^{-1}\,[D,\,e^{\phi}]\,D^{-1}\,[D,\,e^{\phi}]\,D^{-2}.
\]
The first of the two terms vanishes since the residue is a trace. The second
is given by
\begin{equation}
-\,\int\!\!\!\!\!\!-\ \,D^{-1}\,[D,\,e^{\phi}]\,D^{-1}\,[D,\,e^{\phi}%
]\,D^{-2}=\,\int\!\!\!\!\!\!-\ \,[D,\,e^{\phi}]^{2}\,D^{-4}+\,R, \label{com2}%
\end{equation}
where $R$ is the remainder
\[
R=\,-\,\int\!\!\!\!\!\!-\ \,D^{-2}\,[D^{2},\,e^{\phi}]\,D^{-1}\,[D,\,e^{\phi
}]\,D^{-2}.
\]
Let us show that
\begin{equation}
R=-\frac{1}{2}\,\int\!\!\!\!\!\!-\ \,[D^{2},\,e^{\phi}]^{2}\,D^{-6}.
\label{com4}%
\end{equation}
To see this write
\begin{equation}
R=\,-\,\int\!\!\!\!\!\!-\ \,D^{-2}\,[D^{2},\,e^{\phi}]\,D\,D^{-2}%
\,[D,\,e^{\phi}]\,D^{-2}, \label{com5}%
\end{equation}
and note that the commutator of $D$ with $[D^{2},\,e^{\phi}]$ is equal to
$[D^{2},\,[D,e^{\phi}]]$ and has order $1$ so that
\[
\int\!\!\!\!\!\!-\ \,D^{-2}\,[D,[D^{2},\,e^{\phi}]]\,D^{-2}\,[D,\,e^{\phi
}]\,D^{-2}=\,0.
\]
Thus moving $D$ to the left and using the trace property of the residue one
gets
\[
R=\,-\frac{1}{2}\,\int\!\!\!\!\!\!-\ \,[D^{2},\,e^{\phi}]\,D^{-2}%
\,(D\,[D,\,e^{\phi}]+\,[D,\,e^{\phi}]\,D)\,D^{-4},
\]
and one obtains \eqref{com4}. Summarizing, we have shown the equality
\begin{equation}
\int\!\!\!\!\!\!-\ \,(e^{-\phi/2}\,D\,e^{-\phi/2})^{-2}=\,\int
\!\!\!\!\!\!-\ \,e^{2\phi}\,D^{-2}+\,\int\!\!\!\!\!\!-\ \,[D,\,e^{\phi}%
]^{2}\,D^{-4}-\,\frac{1}{2}\,\int\!\!\!\!\!\!-\ \,[D^{2},\,e^{\phi}%
]^{2}\,D^{-6}.\nonumber
\end{equation}
Thus to obtain \eqref{com0} one just needs to prove the equality
\begin{equation}
\int\!\!\!\!\!\!-\ \,[D,\,a]^{2}\,D^{-4}=\,\int\!\!\!\!\!\!-\ \,[D^{2}%
,\,a]^{2}\,D^{-6}, \label{com6}%
\end{equation}
and apply it to $a=e^{\phi}$. One can check \eqref{com6} directly in the
Riemannian case by computing the residue as the integral of the principal
symbols on the unit sphere bundle. The factor $2^{2}$ from the Poisson
brackets $[D^{2},\,a]$ is compensated by the integral of $\xi_{\mu}^{2}$ on
the sphere which gives $\frac{1}{4}$. In the general framework of spectral
triples one gets \eqref{com6} from the general hypothesis
\[
\int\!\!\!\!\!\!-\ \,a\,[D,\,b]\,D^{-3}=\,0\,,\quad\forall a,b\in\mathcal{A}.
\]
Note that the commutator $[D,\,e^{\phi}]$ is skew adjoint and in particular
$[D,\,e^{\phi}]^{\ast}=\,-\,[D,\,e^{\phi}]$. Thus we get the correct sign in
\begin{equation}
\int\!\!\!\!\!\!-\ \,e^{2\phi}\,D^{-2}=\,\int\!\!\!\!\!\!-\ \,(e^{-\phi
/2}\,D\,e^{-\phi/2})^{-2}\,+\,\frac{1}{2}\int\!\!\!\!\!\!-\ \,[D,\,e^{\phi
}][D,\,e^{\phi}]^{\ast}\,D^{-4}. \label{com10}%
\end{equation}

\section{Acknowledgment}

The research of A. Chamseddine is supported in part by the National Science
Foundation under Grant No. Phys-0313416.


\begin{thebibliography}{99}                                                                                               %


\bibitem {Coleman}S. Coleman, \textit{Dilatations }in \textit{Aspects of
Symmetry,} Cambridge University Press, 1985, page 67.

\bibitem {Bud}W. Buchm\"{u}ller and N. Dragon, \textit{Phys. Lett. }\textbf{B
195 }(1987) 417.

\bibitem {Bubu}W. Buchm\"{u}ller and C. Busch, \textit{Nucl. Phys. Proc. Supp.
}\textbf{B 18 }(1990) 295.

\bibitem {Kolb}R. Holman, E. Kolb, S. Vadas and Y. Wang, \textit{Phys. Rev.
}\textbf{D43 }(1991) 3833.

\bibitem {CL}A. Connes and J. Lott, \textit{Nucl. Phys. }\textbf{B 349 }(1991) 71.

\bibitem {CFF}A. H. Chamseddine, G. Felder and J. Fr\"{o}hlich, \textit{Comm.
Math. Phys. }\textbf{155, }109 (1993).

\bibitem {CFG}A. H. Chamseddine, J. Fr\"{o}hlich and O. Grandjean, \textit{J.
Math. Phys. }\textbf{36 }(1995) 6255.

\bibitem {CF}A. Chamseddine and J. Fr\"{o}hlich, \textit{Phys. Lett.
}\textbf{B 314 }(1993) 308.

\bibitem {RS}L. Randall and R. Sundrum, \textit{Phys. Rev. Lett. }\textbf{83
}(1999) 3370.

\bibitem {Lizzi}F. Lizzi, G. Mangano and G. Miele, \textit{Mod. Phys. Lett.
}\textbf{A16 }(2001) 1.

\bibitem {ACAC}A. H. Chamseddine and A. Connes, \textit{Phys. Rev. Lett.
}\textbf{77 }(1996) 4868; \textit{Comm. Math. Phys. }\textbf{186 }(1997) 731.

\bibitem {Kastler}D. Kastler, \textit{Lect. Notes. Phys. }\textbf{543, }(2000) 131-230.

\bibitem {Gilkey}P. Gilkey, \textit{Invariance Theory, the Heat Equation and
the Atiyah-Singer Index Theorem, } Wilmington, Publish or Perish, 1984.

\bibitem {CW}S. Coleman and E. Weinberg, \textit{Phys. Rev. }\textbf{D7,
}(1973) 1888.

\bibitem {Bud2}W. Buchm\"{u}ller and N. Dragon, \textit{Phys. Lett. }\textbf{B
207 }(1988) 292.

\bibitem {van}Y. Ng and H. van Dam, \textit{Phys. Rev. Lett. }\textbf{65
}(1990) 1972.

\bibitem {dragon}W. Buchm\"{u}ller and N. Dragon, \textit{Phys. Lett.
}\textbf{B 223 }(1987) 313.

\bibitem {Ng}Y. Ng, \textit{Int. J. Mod. Phys. }\textbf{D1 }(1992) 145.

\bibitem {Alain}A. Connes, \textit{J. Math. Phys. }\textbf{41 }(2000) 3832.
\end{thebibliography}
\end{document}